\RequirePackage{fixltx2e}
\documentclass[aip,jcp,reprint,floatfix]{revtex4-1}

\usepackage{amsmath,amsfonts} %
\usepackage{wasysym} %
\usepackage[acronym]{glossaries} %
\usepackage{graphicx} %
\usepackage{dcolumn} %
\usepackage{bm} %
\usepackage{ifpdf} %
\usepackage{color} %
\usepackage{nicefrac} %
\usepackage{natbib} %
\usepackage[byname]{smartref} %
\usepackage[breaklinks, %
colorlinks, %
linkcolor=blue, %
citecolor=blue, %
urlcolor=blue]{hyperref} %
\usepackage[table]{xcolor} %
\usepackage{braket}

\newcommand{\KET}[1]{\vert{} #1 \rangle}

\newcommand{\eq}[1]{Eq.~(\ref{#1})}

\newcommand{\fig}[1]{Fig.~\ref{#1}}

\newcommand{\be}{\begin{equation}}
\newcommand{\ee}{\end{equation}}
\newcommand{\bea}{\begin{eqnarray}}
\newcommand{\eea}{\end{eqnarray}}

\ifpdf %
 \fi

\begin{document}
                            
\newacronym{CI}{CI}{conical intersection} %
\newacronym{GP}{GP}{geometric phase} %
\newacronym{TDPT}{TDPT}{time-dependent perturbation theory} %
\newacronym{NAC}{NAC}{non-adiabatic coupling} %
\newacronym{LVC}{LVC}{linear vibronic coupling} %
\newacronym{DOF}{DOF}{degrees of freedom} %
\newacronym{PES}{PES}{potential energy surface} %
\newacronym{BMA}{BMA}{bis(methylene) adamantyl} %
\newacronym{FC}{FC}{Franck-Condon} %
\newacronym{DBOC}{DBOC}{diagonal Born-Oppenheimer correction} %
\newacronym{SH}{SH}{surface hopping} %
\newacronym{EF}{EF}{Ehrenfest} %
\newacronym{EOM}{EOM}{equations of motion}
\newacronym{MQC}{MQC}{mixed quantum-classical}
\newacronym{TDSE}{TDSE}{time-dependent Schr\"odingier equation}

\title{Why do mixed quantum-classical methods describe short-time dynamics through conical intersections so well?
Analysis of geometric phase effects}

\author{Rami Gherib} %
\affiliation{Department of Physical and Environmental Sciences,
  University of Toronto Scarborough, Toronto, Ontario, M1C 1A4,
  Canada} %
\affiliation{Chemical Physics Theory Group, Department of Chemistry,
  University of Toronto, Toronto, Ontario M5S 3H6, Canada} %
\author{Ilya G. Ryabinkin} %
\affiliation{Department of Physical and Environmental Sciences,
  University of Toronto Scarborough, Toronto, Ontario, M1C 1A4,
  Canada} %
\affiliation{Chemical Physics Theory Group, Department of Chemistry,
  University of Toronto, Toronto, Ontario M5S 3H6, Canada} %
\author{Artur F. Izmaylov}
\affiliation{Department of Physical and Environmental Sciences,
  University of Toronto Scarborough, Toronto, Ontario, M1C 1A4,
  Canada} %
\affiliation{Chemical Physics Theory Group, Department of Chemistry,
  University of Toronto, Toronto, Ontario M5S 3H6, Canada} %

\date{\today}

\begin{abstract}
Adequate simulation of non-adiabatic dynamics through conical intersection requires 
account for a non-trivial geometric phase (GP) emerging in electronic and nuclear wave-functions in the adiabatic representation. 
Popular mixed quantum-classical (MQC) methods, surface hopping and Ehrenfest, do not carry a nuclear wave-function to be able
to incorporate the GP into nuclear dynamics. Surprisingly, the MQC methods reproduce ultra-fast interstate crossing dynamics generated with the exact quantum propagation so well as if they contained information about the GP. Using two-dimensional linear vibronic coupling 
models we unravel how the MQC methods can effectively mimic the most significant dynamical GP effects: 
1) compensation for repulsive diagonal second order non-adiabatic couplings and 2) transfer enhancement for a fully cylindrically 
symmetric component of a nuclear distribution. 
\end{abstract}


\maketitle

\glsresetall


\section{Introduction}
\begin{figure}
  \centering
  \includegraphics[width=0.5\textwidth]{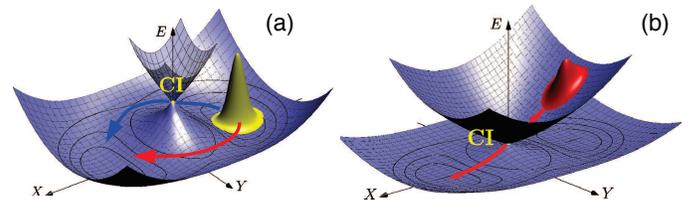}
  \caption{Low energy (a) and excited state (b) nuclear dynamics near \gls{CI}.}
  \label{fig:ci_gp}
\end{figure}

\begin{figure}
  \centering
  \includegraphics[width=0.5\textwidth]{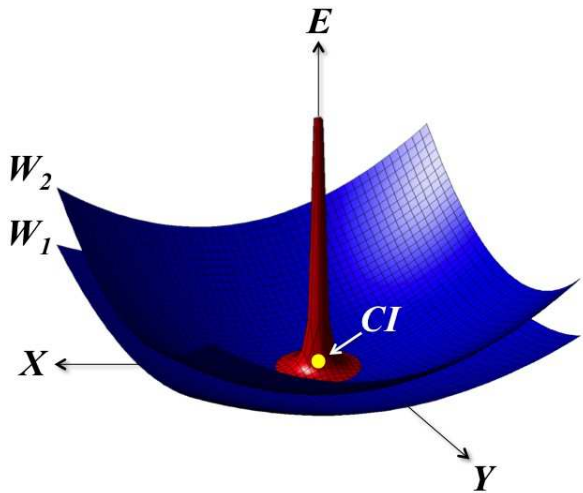}
  \caption{Near CIs, the \gls{DBOC} becomes very large and alters electronic \glspl{PES}. 
  In the absence of the \gls{GP}, the \gls{DBOC} can inhibit the access of a nuclear-wave packet to the \gls{CI}.}
  \label{fig:DBOC}
\end{figure}

The Born-Oppenheimer representation of the electron-nuclear wave-function introduces natural separation between 
time/energy scales of electrons and nuclei in molecular systems. This separation allows one 
to consider nuclear dynamics independently from that of the electronic subsystem reducing the number of involved \gls{DOF}.  
This representation is uniquely defined through the electronic eigenvalue problem with fixed nuclei and 
is conveniently available in numerous electronic structure packages. 
However, there are also a few complications associated with the inherent non-separability of dynamics in a general 
interacting many-body system. For example in many photochemical processes 
nuclear molecular dynamics cannot be adequately modelled on a single \gls{PES} because for some 
nuclear configurations the separation between electronic \glspl{PES} becomes comparable to the 
nuclear energy scale or even vanishes. The latter case often presents itself in the form of \glspl{CI}.\cite{Truhlar:2003/pra/032501,Migani:2004/271}
Non-adiabatic dynamics associated with such crossings not only results in transferring system population 
between electronic states but also in \gls{GP} effects.\cite{Schon:1995/jcp/9292,
Ryabinkin:2013/prl/220406,Baer:1996hl,Ryabinkin:2014/jcp/214116, Althorpe:2008/jcp/214117,Bouakline:2014/cp/} 
The latter is caused by a nontrivial nuclear dependent geometric or Berry phase appearing 
in both nuclear $\chi_j(\mathbf{R},t)$ and electronic $\Phi_j(\mathbf{r};\mathbf{R})$ wave-functions
within the adiabatic representation of the total electron-nuclear wave-function\cite{Cederbaum:2004/CI}
 \bea
 \Psi(\mathbf{r},\mathbf{R},t) &=& \sum_j \Phi_j(\mathbf{r};\mathbf{R})\chi_j(\mathbf{R},t).
 \eea 
 Berry\cite{Berry:1984/rspa/45} and Longuet-Higgins\cite{LonguetHiggins:1958bt} have shown that parametric 
 evolution of the electronic parts $\Phi_j(\mathbf{r};\mathbf{R})$ around the point of eigenvalue 
 degeneracies (the \gls{CI} seam) must change their signs, which 
 makes $\Phi_j(\mathbf{r};\mathbf{R})$ double-valued functions of nuclear \gls{DOF} $\mathbf{R}$.  
 To preserve the single-valued character of the total wave-function,  $\Psi(\mathbf{r},\mathbf{R},t)$, 
 the nuclear part, $\chi_j(\mathbf{R},t)$, must also have 
 a double-valued  character compensating that in the electronic components.  
  \glspl{GP} in electronic wave-functions $\Phi_j(\mathbf{r};\mathbf{R})$ are needed to obtain 
  \glspl{NAC} [$\bra{\Phi_i(\mathbf{R})}\nabla_\mathbf{R} \Phi_j(\mathbf{R})\rangle_\mathbf{r}$ 
  and $\bra{\Phi_i(\mathbf{R})}\nabla_\mathbf{R}^2 \Phi_j(\mathbf{R})\rangle_\mathbf{r}$] 
  correctly,\cite{BenNun:2002tx} but \glspl{NAC} alone are not sufficient for simulating nuclear dynamics properly.
  For the correct dynamics near CIs, the nuclear part must also have the  \gls{GP} resulting in 
  double-valued nuclear wave-functions $\chi_j(\mathbf{R})$. 
  Ignoring the  \gls{GP} of nuclear wave-functions can lead to qualitative distortion of non-adiabatic 
 dynamics even in the the absence of a significant population transfer between 
 crossing electronic states.\cite{Schon:1995/jcp/9292,Baer:1996hl,Ryabinkin:2013/prl/220406,Loic:2013/jcp/234103} 
 Interestingly, dynamical features associated with the  \gls{GP} are very different for low energy dynamics 
 (Fig.~\ref{fig:ci_gp}a) and excited state dynamics (Fig.~\ref{fig:ci_gp}b).
The main manifestation of  \gls{GP} effects in low energy nuclear dynamics is destructive interference between two paths 
around the \gls{CI} seam (Fig.~\ref{fig:ci_gp}a),\cite{Schon:1995/jcp/9292,Ryabinkin:2013/prl/220406,Loic:2013/jcp/234103} 
while in the excited state dynamics (Fig.~\ref{fig:ci_gp}b), it is enhancement of population transfer 
for a fully cylindrical component of a nuclear wave-packet and compensation of a repulsive \gls{DBOC}, 
$\bra{\Phi_i(\mathbf{R})}\nabla_\mathbf{R}^2 \Phi_i(\mathbf{R})\rangle_\mathbf{r}$.\cite{Ryabinkin:2014/jcp/214116}
The \gls{DBOC} is usually neglected in non-adiabatic simulations for molecular systems based on its small value
near the minimum of the ground state. 
However, at the intersecting manifold this term diverges to infinity (see \fig{fig:DBOC}), 
and its {\it a priori} neglect is not justified.  

One of the most popular ways to simulate non-adiabatic dynamics near CIs in large systems is using  
\gls{MQC} approaches: \gls{SH} and \gls{EF} methods.\cite{Tully:1990/jcp/1061,Tully:1998va} 
Unlike full quantum approaches, \gls{MQC} methods propagate nuclear \gls{DOF} classically. 
As a result, they exhibit some well-known limitations of classical mechanics 
such as inability to model nuclear tunnelling and quantum interference effects. 
The  \gls{GP} induced destructive interference in low energy dynamics (Fig.~\ref{fig:ci_gp}a) is a typical example of the latter
effect. \gls{MQC} methods have only the electronic part of the wave-function and thus cannot fully account for
the  \gls{GP}, because even though the electronic function acquires the  \gls{GP} through parametric dependence on nuclear evolution, 
 nuclei evolve according to classical Newton equations that do not have any  \gls{GP} 
 contributions.\footnote{Formally, through the path integral formalism,\cite{Krishna:2007hb} 
 the  \gls{GP} can be introduced into the classical dynamics but it will not change anything because 
 in the classical limit, the \gls{GP} for a \gls{CI} problem results in a delta-function potential at the \gls{CI} point.
 Therefore, the number of classical trajectories influenced by the \gls{GP} is negligible.}
 In this context, it is quite surprising that short-time excited state deactivation dynamics (Fig.~\ref{fig:ci_gp}b) of \gls{MQC} 
 methods agrees extremely well with that of the exact quantum propagation\cite{Muller:1997tq,Worth:2003df,Barbatti:2007cp}
 for systems where  \gls{GP} effects were proven to be very influential.\cite{Ryabinkin:2014/jcp/214116}

In this work we will explain how \gls{MQC} methods emulate \gls{GP} effects in \gls{CI} problems. 
We will restrict our attention to  \gls{GP} effects in excited state deactivation process (Fig. \ref{fig:ci_gp}b). 
As for low energy dynamics (Fig. \ref{fig:ci_gp}a), \gls{SH} and \gls{EF} methods are not much better than 
a simple classical dynamics because non-adiabatic transitions are well suppressed by the energy difference
in areas accessible for classical trajectories. Currently, only the quantum-classical 
Liouville formalism has proven to be capable to capture  \gls{GP} effects in low energy dynamics.\cite{Ryabinkin:2014/jcp/084104}

The rest of the paper is organized as follows. First, we introduce a diabatic 
two-dimensional \gls{LVC} model, although very simple representation 
of the \gls{CI} topology, this model has all quantities involved in quantum and \gls{MQC} simulations
in the analytical form. Thus, it allows us to compare various quantum and \gls{MQC} methods 
on the same footing and to reveal the key components of the \gls{MQC} schemes that are responsible 
for mimicking quantum  \gls{GP} effects. Second, to confirm our analysis
we simulate non-adiabatic dynamics for a few molecular systems that provide 
a variety of dynamical regimes. Finally, we conclude the paper with a summary and further ramifications of our work.


\section{Theory}

\subsection{2D \gls{LVC} model}

The  \gls{GP} appears only in the adiabatic representation, however, it is more convenient to start with a model in the diabatic representation 
because a diabatic model will allow us to obtain the  \gls{GP} explicitly.\cite{Mead:1979/jcp/2284} 
Moreover, the diabatic representation can be exactly transformed to the adiabatic representation 
while the exact reverse transformation does not exist in general.\cite{Mead:1982vm} 
Note that although diabatic wave-functions do not have GPs, simulations in the diabatic representation 
incorporate all  \gls{GP} effects implicitly and are considered to be exact.  
Thus, we start with introducing a diabatic model Hamiltonian
\begin{equation}
  \label{GPLVC2D} 
  H_{\rm 2D}= T_{\rm 2D}{\mathbf 1}_2
  + \begin{pmatrix}
    V_{11} & V_{12} \\
    V_{12} & V_{22}
  \end{pmatrix},
\end{equation}
where $T_{\rm 2D}=-\hbar^2 2^{-1} (\partial^2 /\partial x^2 +\partial^2 /\partial
y^2) $ is the nuclear kinetic energy operator, $x$ and $y$ are
mass-weighted nuclear coordinates, $V_{11}$ and $V_{22}$ are the diabatic
potentials represented by identical two-dimensional parabolas shifted
in space and coupled by the $V_{12}$ potential
\begin{align}
  \label{eq:diab-me-11}
  V_{11} = {} & \dfrac{1}{2}\left[\omega_1^2(x+x_0)^2 + \omega_2^2 y^2\right],
  \quad  V_{12} = c y, \\
  \label{eq:diab-me-22}
  V_{22} = {} & \dfrac{1}{2} \left[\omega_1^2(x-x_0)^2 + \omega_2^2y^2\right].
\end{align}
Here, $\omega_i$'s are harmonic frequencies for nuclear coordinates $x$ and $y$, 
$\pm x_0$ are the minima of $V_{11}$ and $V_{22}$ potentials, and $c$ is a coupling
constant. Electronic \gls{DOF} in $H_{\rm 2D}$ are vectors
$\KET{\phi_1^{\rm dia}}$ and $\KET{\phi_2^{\rm dia}}$ in a two-dimensional linear space. 
To obtain the corresponding adiabatic representation for the Hamiltonian $H_{\rm 2D}$ 
one needs to diagonalize the two-state potential matrix in Eq.~(\ref{GPLVC2D}) by a unitary
transformation that defines the adiabatic electronic states as
\begin{eqnarray}\label{eq:U1}
  \KET{\phi_1^\text{adi}} & = & \phantom{-}\cos\frac{\theta}{2}\,\KET{\phi_1^{\rm dia}} +
  \sin\frac{\theta}{2}\,\KET{\phi_2^{\rm dia}}, \\  \label{eq:U2}
  \KET{\phi_2^\text{adi}} & = & -\sin\frac{\theta}{2}\,\KET{\phi_1^{\rm dia}} +
  \cos\frac{\theta}{2}\,\KET{\phi_2^{\rm dia}},
\end{eqnarray}
where 
\begin{equation}
  \label{eq:theta}
  \theta = \arctan \dfrac{2\,V_{12}}{V_{11} - V_{22}}
\end{equation}
 is a mixing angle between the diabatic states $\KET{\phi_i^{\rm dia}}$.
The adiabatic functions $\KET{\phi_i^\text{adi}}$ are double-valued functions of the nuclear parameters $(x,y)$
because encircling the \gls{CI} point corresponds to change in $\theta(x,y)$ from 0 to 2$\pi$ which leads to a sign 
change in $\KET{\phi_i^\text{adi}}$. 
The unitary transformation to the adiabatic representation brings $H_{\rm 2D}$ to a form
\begin{equation}
  \label{eq:adiab}
  {H}_{\rm 2D}^\text{adi} =   
  \begin{pmatrix}
    T_{\rm 2D} + \tau_{11}& \tau_{12} \\
    \tau_{21} & T_{\rm 2D} +\tau_{22}
  \end{pmatrix} +
  \begin{pmatrix}
    W_{1} & 0 \\
    0 & W_{2}
  \end{pmatrix},
\end{equation}
where
\begin{align}
  \label{eq:W}
  W_{1,2} = & {} \dfrac{1}{2}\left(V_{11} + V_{22}\right) \mp
  \dfrac{1}{2}\sqrt{\left(V_{11} - V_{22}\right)^2 + 4 V_{12}^2},
\end{align}
are the adiabatic potentials with the minus (plus) sign for $W_1$ ($W_2$), and 
\bea\label{eq:taus}
\tau_{ij}&=&-\hbar^2 \mathbf{d}_{ij} \cdot\nabla
- \frac{\hbar^2}{2} D_{ij},
\eea
are kinetic energy terms containing \glspl{NAC}
\bea
  \mathbf{d}_{ij} &=& \bra{\phi_i^\text{adi}} \nabla\phi_j^\text{adi}\rangle, 
  \quad D_{ij} = \bra{\phi_i^\text{adi}} \nabla^2\phi_j^\text{adi}\rangle
  \eea 
with $\nabla = (\partial/\partial x, \partial /\partial y)$. 
Substituting definitions of the adiabatic states into \eq{eq:taus}, $\hat{\tau}_{ij}$ can be express as  
\bea\label{eq:DBOC}
\hat{\tau}_{ii}&=& -\frac{\hbar^2D_{ii}}{2} = \frac{\hbar^2(x^2+y^2)}{8(\gamma^{-1}x^2+\gamma y^2)^2}, \quad \gamma=\frac{c}{x_0\omega_1^2}\\ 
\label{eq:tau12}
\hat{\tau}_{ij}&=&\frac{ \hbar(\overrightarrow{\hat{L}}_z-\overleftarrow{\hat{L}}_z)}{4(\gamma^{-1}x^2+\gamma y^2)}
\label{tau}, \quad i\ne j
\label{eq:gamma}
\eea
where $\hat{L}_z = -i\hbar(x\partial/\partial y - y\partial/\partial x)$ is the angular momentum operator, 
and the overhead arrows indicate operator's directionality.

In quantum dynamics within the adiabatic representation, the \gls{GP} can be introduced
as a position-dependent phase factor $e^{i\theta(x,y)/2}$ for single-valued nuclear basis functions, 
where $\theta(x,y)$ is provided by Eq.~(\ref{eq:theta}).\cite{Mead:1979/jcp/2284} This phase factor would change 
signs of nuclear wave-functions on encircling the \gls{CI}. However, it is more convenient to
use this factor as a unitary transformation of the adiabatic Hamiltonian:  
 $\hat{H}^{\text{GP}}_{2D}=e^{-i\theta/2}\hat{H}^{\text{adi}}_{2D}e^{i\theta/2}$, then 
for simulating \gls{GP} effects we can use a normal single-valued nuclear basis with the $\hat{H}^{\text{GP}}_{2D}$
Hamiltonian.  Also this approach allows one to compare $\hat{H}^{\text{adi}}_{2D}$ and $\hat{H}^{\text{GP}}_{2D}$ 
to see what operator terms are responsible for introducing \gls{GP} effects.
The comparison reveals that the phase factor alters kinetic energy terms
\bea\label{eq:tau11GP}
\hat{\tau}^{\text{GP}}_{ii} &=& \frac{\hbar(\overrightarrow{\hat{L}}_z-\overleftarrow{\hat{L}}_z)}{4(\gamma^{-1}x^2+\gamma y^2)}
+\frac{\hbar^2(x^2+y^2)}{4(\gamma^{-1}x^2+\gamma y^2)^2}, \\ \label{eq:tau12GP}
\hat{\tau}^{\text{GP}}_{ij} &=& \frac{\hbar(\overrightarrow{\hat{L}}_z-\overleftarrow{\hat{L}}_z)}{4(\gamma^{-1}x^2+\gamma y^2)}
-\frac{\hbar^2(x^2+y^2)}{4(\gamma^{-1}x^2+\gamma y^2)^2},~ i\ne j
 \label{taugp}
\eea
and thus changes probabilities of electronic transitions.


\subsection{Mixed quantum-classical methods}


One of the most straightforward routes to the \gls{MQC} methods is to take the classical limit
for nuclear \gls{DOF} in the system wave-function, then for a two-state problem within the adiabatic 
representation an electronic wave-function is given 
\bea
\psi_e (\mathbf{r};\mathbf{R},t) = \sum_{j=1}^2 c_j(t)\phi_j^{\rm adi}(\mathbf{r};\mathbf{R}(t)).
\eea
The time-dependent coefficients are obtained from projection of the 
electronic \gls{TDSE} 
\bea
i\hbar\sum_{j=1}^2 [\dot{c}_j \phi_j^{\rm adi} + c_j \dot{\phi}_j^{\rm adi}] = \sum_{k=1}^2 c_k W_k(\mathbf{R}) \phi_k^{\rm adi}
\eea
onto the electronic adiabatic basis $\phi_j^{\rm adi}(\mathbf{r};\mathbf{R}(t))$ 
\bea\label{eq:cdot}
i\hbar\begin{pmatrix}
{\dot{c}_1} \\
{\dot{c}_2}
  \end{pmatrix} = 
  \begin{pmatrix} 
    W_{1}(\mathbf{R}) & -i\hbar\mathbf{d}_{12}\cdot \dot{\mathbf{R}} \\
    -i\hbar\mathbf{d}_{21}\cdot \dot{\mathbf{R}} & W_{2}(\mathbf{R})
  \end{pmatrix}
 \begin{pmatrix}
{{c}_1} \\
{{c}_2}
  \end{pmatrix}. 
  \eea
Here, the chain rule for \glspl{NAC} is used 
\bea
\bra{\phi_j^{\rm adi}(\mathbf{R}(t))}\dot{\phi}_i^{\rm adi}(\mathbf{R}(t))\rangle = \dot{\mathbf{R}}\cdot \mathbf{d}_{ji}.
\eea 
Thus, due to orthogonality of the adiabatic states the system population of each electronic state 
is given by $|c_i(t)|^2$ and the population transfer is regulated by the off-diagonal 
elements $-i\hbar\mathbf{d}_{12}\cdot \dot{\mathbf{R}}$ in \eq{eq:cdot}.

The nuclear \gls{EOM} for the \gls{EF} method are derived from the total energy conservation condition 
\bea
\frac{dE}{dt} = \frac{d}{dt}\left( \frac{\mathbf{P}^2}{2} + \mathbf{c}^{\dagger}\mathbf{H}_{\rm adi}^{(e)} \mathbf{c} \right)=0,
\eea
with $[\mathbf{H}_{\rm adi}^{(e)}]_{ij} = \delta_{ij} W_i (\mathbf{R})$, which leads to Newton's \gls{EOM} for nuclei
\bea\label{eq:Newt}
\ddot{R}_\alpha = -\mathbf{c}^{\dagger}\frac{\partial{\mathbf{H}_{\rm adi}^{(e)}}}{\partial{R_\alpha}} \mathbf{c} + 
\mathbf{c}^{\dagger}[\mathbf{H}_{\rm adi}^{(e)},\mathbf{d}_{\alpha}] \mathbf{c},
\eea
where 
\bea
\mathbf{d}_{\alpha} &=& 
  \begin{pmatrix} 
    0 & \braket{\phi_1^{\rm adi}\vert\frac{\partial \phi_2^{\rm adi}}{\partial{R_\alpha}}} \\
  \braket{\phi_2^{\rm adi}\vert\frac{\partial \phi_1^{\rm adi}}{\partial{R_\alpha}}} & 0
  \end{pmatrix} 
\eea
Thus, the classical particle moves under an averaged force produced by involved \glspl{PES}. This is even more obvious
if one reformulates the nuclear \gls{EOM} in the diabatic representation
 \bea\label{eq:NewtD}
\ddot{R}_\alpha = -\mathbf{\tilde{c}}^{\dagger}\left[
\frac{\partial}{\partial{R_\alpha}} 
  \begin{pmatrix} 
   V_{11} &  V_{12} \\
   V_{21} & V_{22}
  \end{pmatrix} \right]\mathbf{\tilde{c}}.
\eea 
This reformulation can be done either starting from the beginning using the diabatic representation
\bea
\psi_e (r,t) = \tilde{c}_1(t)\phi_1^{\rm dia}(\mathbf{r})+\tilde{c}_2(t)\phi_2^{\rm dia}(\mathbf{r}),
\eea
or only by applying the adiabatic-to-diabatic transformation 
in \eq{eq:Newt}. This invariance with respect to the electronic state representation 
is one of the advantages of the \gls{EF} method that is not shared by the \gls{SH} method.

In the \gls{SH} case, nuclear \gls{EOM} are also in the Newtonian form but they evolve on a single electronic surface. 
There is some freedom in defining individual electronic surfaces on which nuclear dynamics takes place 
with the only constraint of the energy conservation when a surface switch (hop) takes place.\cite{Tully:1990/jcp/1061}  
This freedom of choice in the electronic surface prompted some works where the \gls{DBOC} had been added 
to the adiabatic states.\cite{Shenvi:2009/jcp/124117,Akimov:2013/jctc/4959} 
The rational can be given if we consider the full quantum nuclear 
equation obtained by projecting the full \gls{TDSE} onto the adiabatic electronic basis
\bea
i\hbar\begin{pmatrix}
{\dot{\chi}_1} \\
{\dot{\chi}_2}
  \end{pmatrix} &=& \Bigg{[}
  \begin{pmatrix}
    T_{\rm 2D} + \tau_{11}& \tau_{12} \\
    \tau_{21} & T_{\rm 2D} +\tau_{22}
  \end{pmatrix} \\ 
  &&+ 
  \begin{pmatrix}
    W_{1} & 0 \\
    0 & W_{2}
  \end{pmatrix}\Bigg{]}
 \begin{pmatrix}
{{\chi}_1} \\
{{\chi}_2}
  \end{pmatrix}. 
  \eea 
  Grouping all diagonal potential-like terms involves the second order \glspl{NAC} 
  $-\hbar^2 D_{ii}/2$ which are functions of $\mathbf{R}$, therefore, they 
  can be added to the potential energy surfaces $W_{i}$
  to create modified surfaces 
  \bea\label{eq:Wt}
  \widetilde{W}_{i}(\mathbf{R}) = W_{i}(\mathbf{R})- \frac{\hbar^2}{2}D_{ii}(\mathbf{R}). 
  \eea
  Considering that the \gls{DBOC} has the $\hbar^2$ prefactor, its addition may seem insignificant. 
  However, $D_{ii}(\mathbf{R})$ diverges at the \gls{CI} point [\eq{eq:DBOC}, and \fig{fig:DBOC}], and hence, 
  in the vicinity of the \gls{CI} the \gls{DBOC} cannot be neglected based on its relatively small values far from the \gls{CI}. 
  
Note that there are no $\hbar$ terms present in the force definition in \eq{eq:Newt}, this shows completely classical 
nature of the nuclear \gls{EOM}. Also, even though we work in the adiabatic representation, the second order  
derivative couplings do not appear in the working equations because the nuclear kinetic energy has not been 
considered as a quantum operator. Besides the reason of inconsistency in powers of $\hbar$, 
introducing the \gls{DBOC} into the \gls{EF} method would break the invariance of this method 
with respect to the electronic state representation.  

\subsection{Mexican hat model}

The adiabatic potentials $W_{i}$ in \eq{eq:W} of the 2D \gls{LVC} Hamiltonian 
acquires a cylindrical symmetry for $\gamma=1$ [\eq{eq:DBOC}]. This symmetry facilitates comparison of non-adiabatic 
transfer elements for different methods. Thus, we will consider in details the 
$\gamma=1$ case, also known as the Mexican hat model. 


For our analysis, it is convenient to write $H_{\rm 2D}$ [\eq{GPLVC2D}]
with $\omega_1=\omega_2=x_0=1$ and $\Delta=0$
in polar $(\rho, \varphi)$ coordinates centered at the \gls{CI}:
\begin{align}
  \label{eq:mhHam}
  \hat H_M & = \frac{1}{2}\left(-\frac{\hbar^2}{\rho}\frac{\partial}{\partial \rho}
    \rho \frac{\partial}{\partial \rho} + \frac{\hat L_z^2}{\rho^2} + \rho^2
  \right) \mathbf{1}_2 \nonumber \\
  & + \rho
  \begin{pmatrix}
    \cos\varphi & \sin\varphi \\
    \sin\varphi & -\cos\varphi
  \end{pmatrix},
\end{align}
where $\hat{L}_z = -i\hbar\partial/\partial \varphi$.
It is easy to verify that $\hat H_M$
commutes with the vibronic angular momentum operator $\hat J = \hat
L_z\mathbf{1}_2 + \frac{\hbar}{2}\boldsymbol\sigma_y$, where $\boldsymbol\sigma_y$ is the Pauli
matrix. Eigenfunctions of $\hat J$ are
\bea
  \label{eq:Jdia_eig}
  \langle \varphi\ket{m}_1^{\rm dia} = e^{im\varphi} 
  \begin{pmatrix}
    \cos\frac{\varphi}{2} \\
    \sin\frac{\varphi}{2}
  \end{pmatrix},
  \eea
  and
  \bea 
  \label{eq:Jdia_eig2}
  \langle \varphi\ket{m}_2^{\rm dia} = e^{im\varphi}
  \begin{pmatrix}
    -\sin\frac{\varphi}{2} \\
    \cos\frac{\varphi}{2}
  \end{pmatrix},
\eea
where $m$ are half-integer eigenvalues.
Functions $\langle \varphi \ket{m}_i^{\rm dia}$ are single-valued as eigenfunctions of a 
general quantum-mechanical operator without external parameter dependence should be. 

Let us now transform both operators to the adiabatic
representation. For this model, the $\theta$ angle [\eq{eq:theta}] of the unitary transformation [Eqs.~\eqref{eq:U1} and \eqref{eq:U2}] 
becomes the polar angle $\varphi$. The unitary transformation of $\hat H_M$ and ${\hat J}$ leads to 
\bea \notag
  \hat H_M^\text{adi} &=& \frac{1}{2}\left(-\frac{\hbar^2}{\rho}\frac{\partial}{\partial \rho} \rho
    \frac{\partial}{\partial \rho} + \rho^2 \right) {\mathbf 1}_2  \\ \label{eq:Hmh_adi}
    &&+ \frac{\left(\hat L_z{\mathbf 1}_2 -\frac{\hbar}{2}\boldsymbol\sigma_y\right)^2}{2\rho^2}  + \rho
    \boldsymbol\sigma_z, \\
  \label{eq:Jadi}
  \hat J^\text{adi} &=& \hat L_z{\mathbf 1}_2.
\eea
The transformation of the $\hat J$ eigenfunctions [Eqs.~\eqref{eq:Jdia_eig} and \eqref{eq:Jdia_eig2}]
 gives $ \langle \varphi\ket{m}_1^\text{adi} = (e^{im\varphi}, 0)^\dagger$
and $ \langle \varphi\ket{m}_2^\text{adi} = (0, e^{im\varphi})^\dagger$, which seem as
regular eigenfunctions of $\hat L_z$ apart from their half-integer values of $m$. 
Thus, the eigenfunctions of $\hat
J^\text{adi}$ are \emph{double-valued} functions. 

Since the commutation relations are the same in all representations,
$\hat H_M^\text{adi}$ and $\hat J^\text{adi}$ commute and have a common
system of eigenfunctions, hence, the eigenfunctions of $\hat H_M^\text{adi}$ [\eq{eq:Hmh_adi}] can be sought as
\begin{equation}
  \label{eq:model_eig}
  \begin{pmatrix}
    \langle \rho,\varphi\ket{\chi_1} \\
    \langle \rho,\varphi\ket{\chi_2}
  \end{pmatrix} 
  = 
  \begin{pmatrix}
    \langle \rho\ket{f_1} \\
    \langle \rho\ket{f_2} 
  \end{pmatrix}\langle \varphi\ket{m}.
\end{equation}
In Eq.~(\ref{eq:model_eig}) the double-valuedness of adiabatic nuclear
wave-functions is isolated in $\langle \varphi\ket{m}=e^{im\varphi}$. 
One can turn \gls{GP} effects ``on'' and ``off'' by changing $m$: half-integer
values correspond to inclusion of the \gls{GP}, whereas integer values
mean the \gls{GP} is neglected. 

\begin{figure}[!h]
  \centering
  \includegraphics[width=0.5\textwidth]{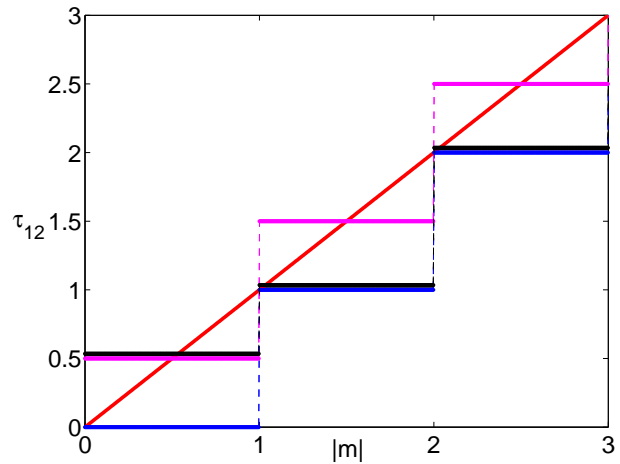}
  \caption{Transition amplitude $\tau_{12}$ as a function of $|m|$ for different methods: 
  quantum without the GP \eq{eq:tau_nogp} (blue), 
  quantum with the GP \eq{eq:tau_gp} (black), continuous classical 
  \eq{eq:tau_cc} (red), binned classical \eq{eq:tau_cl} (magenta). $\rho$-dependent parts are neglected in all cases.}
  \label{fig:steps}
\end{figure}

To perform comparative analysis of quantum methods with and without \gls{GP} with the same 
set of integer $m$ angular functions, we apply a gauge transformation of 
Mead and Truhlar,\cite{Mead:1979/jcp/2284}  $e^{i\varphi/2}$, to all half-integer angular functions $\ket{m}$. 
This transformation changes the $\hat H_M^\text{adi}$ Hamiltonian by modifying kinetic energy terms
as in the case of the general 2D \gls{LVC} model [Eqs.~\eqref{eq:tau12GP} and \eqref{eq:tau11GP}]. 
For the Mexican hat model we can separate angular and radial components for all kinetic energy terms 
which allows for more detailed analysis. To estimate $m$ dependence of quantum transition probabilities without breaking 
symmetry with respect to left and right rotation around the \gls{CI} point we consider sums of $\tau_{12}$ elements 
projected onto $\ket{\pm m}\ket{f_i}$ states. For the Mexican hat Hamiltonian without explicit \gls{GP} terms [\eq{eq:Hmh_adi}]
transition amplitudes are
\bea\notag
\tau_{12} (|m|) &=& \frac{\hbar}{2}\left\vert \bra{f_1}\bra{m} \frac{\hat{L}_z}{2\rho^2}\ket{m}\ket{f_2}\right\vert \\
&+& \frac{\hbar}{2}\left\vert  \bra{f_1}\bra{-m} \frac{\hat{L}_z}{2\rho^2} \ket{-m}\ket{f_2}\right\vert  \\ \label{eq:tau_nogp}
&=& \hbar |m| \left\vert \bra{f_1} \frac{1}{2\rho^2}\ket{f_2}\right\vert.
\eea
Adding the \gls{GP} modifies transition elements as  
\bea\notag
\tau_{12}^{\rm GP}(|m|) &=&  \frac{\hbar}{2}\left\vert \bra{f_1}\bra{m} \frac{\hat{L}_z-\hbar/2}{2\rho^2}\ket{m}\ket{f_2}\right\vert \\
&+& \frac{\hbar}{2}\left\vert \bra{f_1}\bra{-m}  \frac{\hat{L}_z-\hbar/2}{2\rho^2}\ket{-m}\ket{f_2}\right\vert  \\
&=& \frac{|m-\frac{1}{2}| +|m+\frac{1}{2}|}{2} \left\vert \bra{f_1} \frac{\hbar^2}{2\rho^2}\ket{f_2}\right\vert \\ \label{eq:tau_gp}
&=& \left\{
\begin{array}{c l}     
    \frac{\hbar^2}{2}\left\vert \bra{f_1} \frac{1}{2\rho^2}\ket{f_2}\right\vert, & |m|=0\\
    \hbar^2 |m| \left\vert \bra{f_1} \frac{1}{2\rho^2}\ket{f_2}\right\vert, & |m|\ne0
\end{array}\right.
\eea
Note that the only difference from adding the \gls{GP} correction is a transition enhancement for the $m=0$ component in 
the presence of the \gls{GP} (\fig{fig:steps}). 

For the classical treatment of nuclear motion in \gls{MQC} methods, we note that the classical nuclear angular momentum 
is conserved because of the cylindrical symmetry. The transition amplitude is given by  
\bea\label{eq:tau_cc}
\tau_{12}^{\rm MQC} = i\hbar\mathbf{d}_{21}\cdot \dot{\mathbf{R}} = \frac{L_z}{2\rho^2},
\eea
where $L_z=xP_y-yP_x$ is the classical angular momentum.
For comparison with the quantum results we will perform quasi-classical binning by integrating 
continuous ${L_z=\hbar m}$ values between discrete values of $m$ and $m+1$ 
\bea\label{eq:tau_cl}
\tau_{12,\rm b}^{\rm MQC} (m) = \hbar\int_{m}^{m+1}\frac{mdm}{2\rho^2} =  \frac{\hbar(m+1/2)}{2\rho^2}.
\eea
The same contribution will appear if we consider the $[-m,-m-1]$ range, thus the averaging of two results does not change the outcome
$\tau_{12,\rm b}^{\rm MQC} (|m|) = |\tau_{12,\rm b}^{\rm MQC} (m)|/2 + |\tau_{12,\rm b}^{\rm MQC} (-m)|/2 = \tau_{12,\rm b}^{\rm MQC} (m)$.
If one neglects the difference between the classical $1/\rho^2$ term and its quantum analogue, then for $m=0$
$\tau_{12,\rm b}^{\rm MQC}$ and $\tau_{12}^{\rm GP}$ are the same (\fig{fig:steps}).  This is a result of 
a continuous nature of the classical angular momentum that after integrating over the range $|m|\in[0,1]$
introduces a \gls{GP}-like enhancement. 

\begin{figure}[!h]
  \centering
  \includegraphics[width=0.5\textwidth]{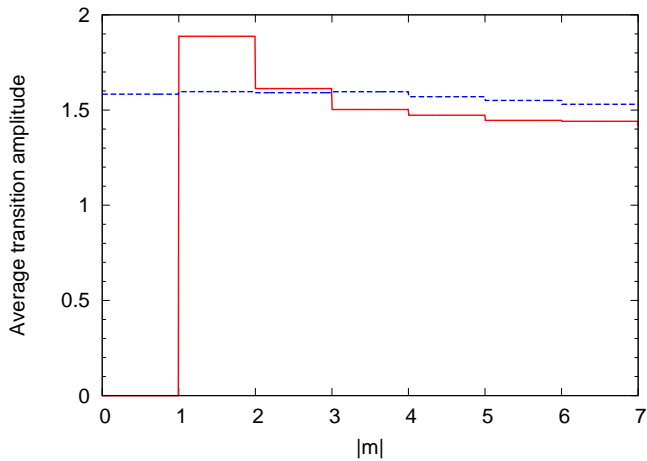}
  \caption{Transition amplitudes, $\hbar^2|m\bra{f_1}(2\rho)^{-2} \ket{f_2}|$ for quantum (sold red) 
  and $\hbar(|m|+1/2) (2\rho)^{-2}$ for MQC (dashed blue) simulations, 
  averaged over the first a.u. of simulation time as a function of $|m|$. }
  \label{fig:transprob}
\end{figure}

However, the solely angular dependence consideration raises a question whether the continuous character of the 
classical angular momentum will cause overestimation of the transition probability for high $|m|$'s (\fig{fig:steps}, ${|m|>1}$). 
The answer is negative, and it becomes obvious if we account for the $\rho$-dependence of $\tau_{12}$ terms. 
Although $\rho$ and $L_z$ are independent, intuitively, 
it is clear that for a general trajectory, due to the centrifugal force, high $m$'s have low $1/\rho^2$ factors in \eq{eq:tau_cl}.
Therefore, we can approximate ${(m+1/2)/\rho^2\sim m/\rho^2}$ for large $|m|$'s. To confirm this intuitive consideration we 
performed both classical and quantum simulations for the Mexican hat model. In quantum simulations 
the initial state is given by a Gaussian wave packet placed on the upper cone
\begin{equation}
  \label{eq:gwp}
  \chi_2(x,y,0) = \sqrt{\frac{2}{\pi \sigma_x \sigma_y}}
  \exp{\left(-\frac{(x-x_0)^2}{\sigma_x^2} -
      \frac{y^2}{\sigma_y^2}\right)}
\end{equation} 
with widths $\sigma_x =\sigma_y= \sqrt{2}$. The classical counterpart 
is initiated from a Gaussian distribution for positions and momenta obtained via the Wigner transform of $\chi_2(x,y,0)$. 
Average transition amplitudes split to $m$ components in \fig{fig:transprob} confirm the conjecture of 
the radial component ($1/\rho^2$) reduction with increase of the angular component ($m$). Moreover,
these effects compensate each other consistently through the $m$ series so that both methods plateaux at large $|m|$'s.
Therefore, a continuous character of classical angular momentum helps to mimic the enhancement of the fully 
cylindrical $m=0$ component and does not interfere with other angular components.  

\section{Molecular calculations}

Here we consider three molecular systems: the bis(methylene) adamantyl cation, 
(BMA)\cite{Blancafort:2005/jacs/3391,Izmaylov:2011/jcp/234106}
the butatriene cation,\cite{Koppel:1984/acp/59,Cederbaum:1977/cp/169,Cattarius:2001/jcp/2088,Sardar:2008/pccp/6388,
Burghardt:2006/mp/1081,Gindensperger:2006/jcp/144103} 
and the pyrazine molecule.\cite{Seidner:1993/cpl/117,Woywod:1994/jcp/1400,Sukharev:2005/pra/012509}
These systems have been extensively studied before, and it was shown that they are well  
described by multi-dimensional \gls{LVC} models. For our simulations, we have reduced N-dimensional \gls{LVC} models to 
effective 2D \gls{LVC} models using collective nuclear DOF so that the 2D models reproduce short-term dynamics 
of the original N-dimensional models [see the Appendix of Ref.~\onlinecite{Ryabinkin:2014/jcp/214116} for details].
These three systems have \gls{LVC} parameters (Table~\ref{tab:BMA-param}) that are representative for various 
manifestations of  \gls{GP} effects.\cite{Ryabinkin:2014/jcp/214116}
\begin{table}
  \caption{Parameters of the 2D effective \protect\gls{LVC}
    Hamiltonian, Eq.~\eqref{GPLVC2D} for the three studied systems.} 
  \label{tab:BMA-param}
  \centering
  \begin{ruledtabular}
    \begin{tabular}{@{}lccccr@{}}
      \multicolumn{1}{c}{$\omega_1$} & $\omega_2$ & $a$ & $c$ &
      \multicolumn{1}{c}{$\Delta$} & $\gamma$ \\ \hline
      \multicolumn{6}{c}{ Bis(methylene) adamantyl cation} \\
      $7.743\times10^{-3}$ & $6.680\times10^{-3}$ & 31.05 &
      $8.092\times 10^{-5}$ & 0.000 & 0.09 \\[1ex]
      \multicolumn{6}{c}{ Butatriene cation} \\
      $9.557\times10^{-3}$ & $3.3515\times10^{-3}$  & 20.07   &
      $6.127\times 10^{-4}$ & 0.020 & 0.67 \\[1ex]
      \multicolumn{6}{c}{ Pyrazine} \\
      $3.650\times10^{-3}$ & $4.186\times10^{-3}$ & 48.45 & $4.946\times
      10^{-4}$ & 0.028 & 1.50
    \end{tabular}
  \end{ruledtabular}
\end{table}

To provide comparative analysis of dynamical features appearing from \gls{DBOC} and GP influences
in addition to \gls{MQC} simulations we provide quantum results obtained using three nuclear 
non-adiabatic Hamiltonians:
1) the full Hamiltonian [\eq{GPLVC2D}] producing {\it exact} dynamics, 2) the ``no  \gls{GP}'' Hamiltonian 
[Eqs.~\eqref{eq:adiab}, \eqref{eq:DBOC}, and \eqref{eq:tau12}],
and 3) the ``no GP, no \gls{DBOC}'' Hamiltonian [Eqs.~\eqref{eq:adiab}, \eqref{eq:tau12}, and $\tau_{ii}=0$].
Numerical procedures to propagate the \gls{TDSE} with 
these Hamiltonians are detailed in Ref.~\onlinecite{Ryabinkin:2014/jcp/214116}.
For all methods we compare the adiabatic population of the excited electronic state 
$P_{\rm adi}^{(e)}(t)  =  \braket{\chi_{2}(t)|\chi_2(t)}$,
where $\chi_2(x,y,t)$ is a time-dependent nuclear wave-function
that corresponds to the excited adiabatic electronic state initiated as  
a Gaussian distribution in \eq{eq:gwp} with widths $\sigma_x =\sqrt{2/\omega_1}$ and $\sigma_y = \sqrt{2/\omega_2}$. 
To calculate $P_{\rm adi}^{(e)}(t)$ in \gls{MQC} simulations, $|c_2|^2(t)$ are used in the \gls{EF} approach, 
and the instantaneous ratios between the number of trajectories on the excited state to the total number of trajectories 
are taken in the \gls{SH} approach. In both MQC methods, $P_{\rm adi}^{(e)}(t)$ are averaged over 2000 trajectories 
with nuclear momenta and positions sampled from the Wigner transform of 
the corresponding quantum wave packet. To integrate MQC \gls{EOM}, the 4th order Runge-Kutta integrator 
has been used with the time-step 0.05 fs for \gls{SH} and 0.001 fs for \gls{EF} methods, respectively.  
For the \gls{SH} method we used Tully's fewest switches algorithm\cite{Tully:1990/jcp/1061} 
with nuclear forces obtained from adiabatic \glspl{PES}, $W_i$ [\eq{eq:W}]. To illustrate the influence of the DBOC in 
\gls{SH} dynamics, we introduce a modification, further referred as SH+DBOC, where 
nuclear forces are obtained from adiabatic \glspl{PES} with 
the DBOC, $\tilde{W}_i$ [\eq{eq:Wt}].   

As has been established in our previous study,\cite{Ryabinkin:2014/jcp/214116}  
\gls{GP} effects manifest themselves quite differently in 
these molecular models: for BMA,  \gls{GP} effects are 
predominantly in compensation of a potential repulsion introduced by the \gls{DBOC}, while for the butatriene cation
and the pyrazine molecule the \gls{GP} enhances non-adiabatic transfer of wave-packet's $m=0$ component. 
This difference stems from the anisotropy of the branching space, which is well characterized by the value of 
$|\gamma-\gamma^{-1}|$, the smaller this value is the closer the problem to the Mexican hat case is and more cylindrical 
all potential terms including the \gls{DBOC} are. For BMA, $|\gamma-\gamma^{-1}|=11.4$ and this makes the \gls{DBOC} a wide 
repulsive wall, while for the other systems, $|\gamma-\gamma^{-1}|\sim 0.8$ which is much closer to 
the Mexican hat limit ($|\gamma-\gamma^{-1}|=0$).  
Thus, we will initially analyze performance of the \gls{MQC} methods for the BMA cation 
and then for the other two systems.

\begin{figure}
  \centering
  \includegraphics[width=0.5\textwidth]{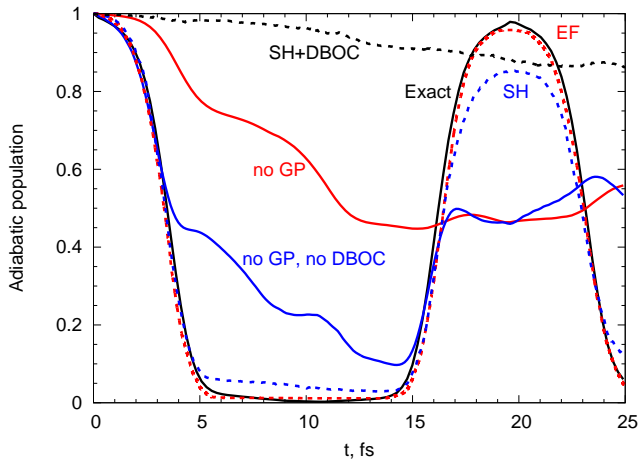}
  \caption{Excited state adiabatic population ($P_{\rm adi}^{(e)}$) dynamics for the BMA cation obtained with different methods.}
  \label{fig:BMA}
\end{figure}

\paragraph{BMA: \gls{DBOC} in MQC.}
The exact population dynamics in BMA demonstrates coherent oscillations that can be easily 
understood considering weak diabatic couplings in this system (\fig{fig:BMA}). Thus, the exact dynamics 
almost solely undergoes on a single diabatic state that corresponds to the excited adiabatic state
before the crossing and the ground adiabatic state after the crossing. Excited state populations in both 
\gls{MQC} methods reproduce almost exactly those of the full QM dynamics. 
%
%
Smaller amplitudes of adiabatic population oscillations in \gls{SH} than those in the exact dynamics
is a manifestation of \gls{SH} overestimation of diabatic population transfer. The origin of this 
overestimation and violation of the Marcus theory limit
has been found in higher electronic coherences within the \gls{SH} approach.\cite{Landry:2011ej,Landry:2012hr} 
The explanation of the overall success of the MQC methods for BMA
is in the absence of both  \gls{GP} and \gls{DBOC} terms in these methods.
Therefore the \gls{MQC} methods do not need the \gls{GP} for cases where the main role of 
the \gls{GP} is the \gls{DBOC} compensation. On the other hand, comparing the SH+DBOC dynamics
with the exact one shows that adding the \gls{DBOC} 
to the adiabatic potentials can be very detrimental for \gls{MQC} results (\fig{fig:BMA}). The impact of uncompensated 
\gls{DBOC} terms in \gls{MQC} dynamics is even bigger than removing \gls{GP} terms in 
quantum simulations. This is a result of a repulsive potential of the \gls{DBOC} that prevents classical 
nuclear dynamics in the \gls{SH} method to approach a region of strong non-adiabatic coupling.
Quantum wave packets for the ``no GP''  Hamiltonian can increase the non-adiabatic transfer due to some
tunnelling under the \gls{DBOC} potential.  
Thus these results demonstrate that the \gls{DBOC} should not be added to the \gls{MQC} methods.

Although the BMA branching space is significantly anisotropic, 
the transfer of the wave-packet $m=0$ component is still somewhat suppressed.\cite{Ryabinkin:2014/jcp/214116}
Since the weight of the wave-packet $m=0$ component near the CI for BMA is quite substantial, 42\%, 
absence of the GP enhancement of its transition in the ``no DBOC, no GP'' model results in deviation of the corresponding  
population dynamics from the exact one after 4 fs. 
Thus, even in anisotropic systems the GP induced non-adiabatic transfer 
enhancement and its imitation by MQC methods can be crucial for 
quantitative agreement with exact dynamics.

\begin{figure}[!h]
  \centering
  \includegraphics[width=0.5\textwidth]{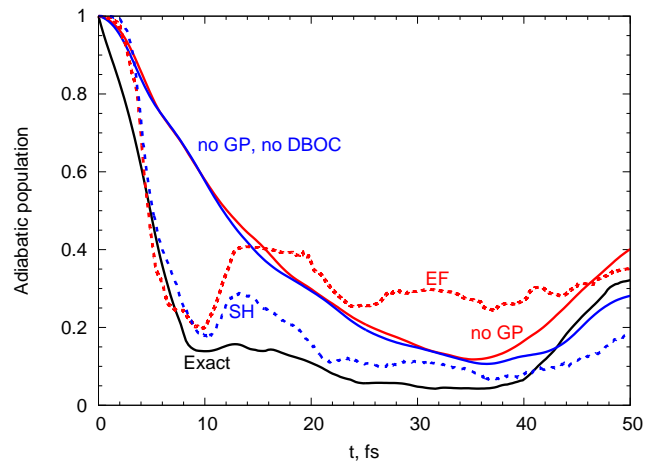}
  \caption{Excited state adiabatic population ($P_{\rm adi}^{(e)}$) dynamics of the butatriene cation obtained with different methods.}  
  \label{buta}
\end{figure}

\begin{figure}[!h]
  \centering
  \includegraphics[width=0.5\textwidth]{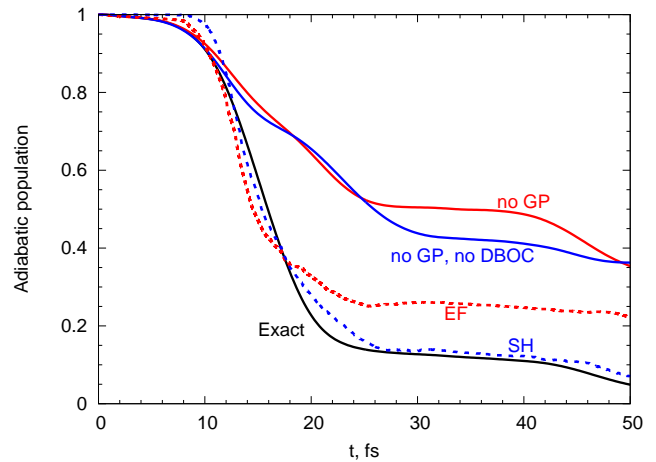}
  \caption{Excited state adiabatic population ($P_{\rm adi}^{(e)}$) dynamics of pyrazine obtained with different methods.}  
  \label{pyra}
\end{figure}

\paragraph{C$_4$H$_4^{+}$ and pyrazine: $m=0$ enhancement in \gls{MQC}.}
In the butatriene cation and the pyrazine molecule, the \gls{DBOC} potential is 
relatively isotropic and compact.  Therefore, it does not prevent a nuclear wave packet 
from accessing the vicinity of the CI, and the \gls{DBOC} presence does not change quantum 
dynamics significantly (see Figs.~\ref{buta} and \ref{pyra}). Yet, \gls{GP} effects are significant, because the \gls{GP} related 
term in $\tau_{12}^{\rm GP}$ accelerates the non-adiabatic transfer for the $m=0$ component. 
Interestingly, this acceleration is well mimicked by the \gls{MQC} methods due to the classical 
description of the angular momentum which leads to similar enhancement of the $m=0$ component
transfer. According to the angular decomposition of quantum wave packets at the moment of the closest
proximity to the \gls{CI}, in both systems, the weight of the $m=0$ component is close to 90\%. 
Therefore, this enhancement is the main  \gls{GP} effect in these systems. 
The excited state population dynamics in Figs.~\ref{buta} and \ref{pyra} reveal 
that the \gls{MQC} methods can reproduce the exact quantum dynamics and perform 
better than quantum methods that do not account for the \gls{GP}.
In both systems, the \gls{SH} method performs slightly better compared to the \gls{EF} approach.

\section{Conclusions}

It has been recently found that pure quantum effects associated with the nontrivial geometric phase
appearing in the nuclear and electronic adiabatic wave-functions for surface crossing problems
can significantly affect population transfer dynamics. Although \gls{MQC} methods ignore nuclear \gls{GP} 
effects by substituting quantum nuclear dynamics with its classical approximation, they are still very successful 
in simulating non-adiabatic dynamics through \glspl{CI}. In this work we have 
unraveled the key elements of this success: 
Both types of  \gls{GP} effects involved in the excited state dynamics, 
the \gls{DBOC} compensation and the enhancement of the non-adiabatic transfer for the 
fully cylindrical component of a wave-packet, are mimicked fortuitously in \gls{MQC}
methods using classical mechanics. 

Interestingly, the \gls{DBOC} term did not appear in original derivations of \gls{MQC} 
schemes and have been added only later in an {\it ad hoc} manner. This work clearly demonstrated that 
such addition can be very detrimental for the quality of results and should be avoided.
The mechanism for the cylindrical component enhancement in the \gls{MQC} schemes 
has been elaborated on the Mexican hat model. The situation in some sense is opposite to the 
famous Planck quantization via discrete summation to describe the black-body radiation.  
In \gls{MQC} transfer element, the purely quantum effect from the  \gls{GP} is recovered because a discrete 
summation over the angular eigenstates of the angular momentum operator is substituted by a classical 
continuous integration. Thus, nuclear \gls{GP} effects make excited state 
non-adiabatic dynamics more classical by compensating some other quantum effects.

There have been several proposals on using the Landau-Zener (LZ) formula\cite{Landau:Z,Zener:1932wr} to model non-adiabatic 
dynamics through the conical intersection.\cite{Teller:1937vi,Alijah:1999ic,Belyaev:2014cb,Malhado:2008ia} 
These proposals involve application of the LZ equation for probability transfer with a subsequent averaging over individual classical trajectories. 
Surprisingly, a question of influence of the conical intersection topology on the result has not ever been raised. 
The current work can be used to rationalize an application of LZ-based approaches to the \gls{CI} problem even though 
in such methods topological geometric phase effects are not explicitly accounted. 

\section{Acknowledgements}

A.F.I. acknowledges stimulating discussion with John Tully and 
funding from the Natural Sciences and Engineering Research Council of Canada (NSERC) through
the Discovery Grants Program. R.G. is grateful to the Chemistry Department of the University of Toronto
for a summer research fellowship for newly admitted graduate students.

%

\end{document}